\documentclass[twocolumn,showpacs,prl,amsmath,amssymb]{revtex4}
\usepackage{graphicx}
\usepackage{dcolumn}
\usepackage{bm}

\begin{document}

\title{First principles calculation of vibrational Raman spectra in large 
systems: signature of small rings in crystalline SiO$_2$}
\author{Michele Lazzeri and Francesco Mauri}
\affiliation{Laboratoire de Min\'eralogie Cristallographie de Paris,
4 Place Jussieu, 75252 Paris cedex 05, France.}
\date{\today}

\begin{abstract}
We present an approach for the efficient
calculation of vibrational Raman intensities in periodic systems within
density functional theory.
The Raman intensities are computed from the second order derivative of the
electronic density matrix with respect to a uniform electric field.
In contrast to previous approaches, the computational effort required by our 
method for the evaluation of the intensities is negligible compared to that 
required for the calculation of vibrational frequencies.
As a first application, we study the signature of 3- and 4-membered rings in the the Raman spectra 
of several polymorphs of SiO$_2$, including a zeolite having 102 atoms per unit cell.
~\\
\end{abstract}



\maketitle

Vibrational Raman spectroscopy~\cite{bruesch} is one of the most widely used
optical techniques in materials science.
It is a standard method for quality control in production lines.
It is very effective in determining the occurrence of new
phases or structural changes at extreme conditions (high pressure and 
temperature), where it is often preferred to the more difficult and
less readily available
x-ray diffraction experiments based on synchrotron sources
~\cite{hemley87}.
Moreover, it can be used in the absence of long-range structural order
as for liquid or amorphous materials~\cite{sharma81,hosono01,pasquarello98}.
The theoretical determination of Raman spectra 
is highly desirable, since it can be used to associate Raman 
lines to specific microscopic structures.

Density functional theory (DFT)~\cite{DFT} can be used to determine with
high accuracy both frequencies and intensities of Raman spectra.
Vibrational frequencies can be efficiently determined using
{\it first order response}~\cite{DFPT,gonze95}.
Within this approach Raman intensities (RI) calculation is also
possible, but requires a computational time significantly larger
and is not practical for large systems.
Thus, while many examples of frequency calculations have been reported so far
~\cite{DFPT}, RI were predicted from first-principles in a very limited
number of cases involving systems with a small number of atoms
~\cite{baroni86,umari01,putrino02}.
In this Letter we show that it is possible to obtain RI in extended solids with 
a computational
cost negligible with respect to that required for the frequency determination.
The efficiency of our approach will lead {\it ab-initio} calculations
to become a routine instrument for the interpretation of experimental
Raman data.
Our method is based on {\it second order response} to DFT.
In particular, we compute the second order derivative of
the electronic density matrix with respect to a uniform electric field,
using pseudopotentials and periodic boundary conditions.
As a first application we calculate
Raman spectra of several SiO$_2$ polymorphs, including a zeolite
having 102 atoms per unit cell~\cite{zeolite}.

In a Raman spectrum the peak positions are fixed by the frequencies  
$\omega_\nu$ of the optical phonons with null wavevector.
In non-resonant Stokes Raman spectra of harmonic solids, the peak intensities $I^\nu$ can be computed 
within  the Placzek approximation~\cite{bruesch} as:
\begin{equation}
\label{eq1}
I^\nu\propto |{\bf e}_i\cdot\tensor{\bf A}^{\nu}\cdot{\bf e}_s|^2
\frac{1}{\omega_{\nu}}(n_\nu+1),
\end{equation}
where ${\bf e}_i$ (${\bf e}_s$) is the polarization of the
incident (scattered) radiation, $n_\nu=(\exp(\hbar\omega_\nu/k_{\rm B}T)-1)^{-1}$,
$T$ is the temperature, and
\begin{equation}
\label{eq2}
A^\nu_{lm}=\sum_{k\gamma}
\frac{\partial^3{\cal E}^{\rm el}}{\partial E_l \partial E_m
\partial u_{k\gamma}}
\frac{w^\nu_{k\gamma}}{\sqrt{M_{\gamma}}}.
\end{equation}
Here ${\cal E}^{\rm el}$ is the electronic energy of the system,
$E_l$ is the $l^{th}$ Cartesian component
of a uniform electric field,
$u_{k\gamma}$ is the displacement of the $\gamma^{th}$ atom in the
$k^{th}$ direction, $M_{\gamma}$ is the atomic mass, and
$w^\nu_{k\gamma}$ is the orthonormal vibrational eigenmode $\nu$.

Linear response
~\cite{DFPT,gonze95} can be used to determine $\omega_\nu$, ${\bf w}^\nu$,
and also the dielectric tensor $\tensor{\bm{\epsilon}}^\infty$ defined as
$\epsilon^\infty_{lm}=\delta_{lm}-(4\pi/{\Omega})
\partial^2{\cal E}^{\rm el}/({\partial E_l \partial E_m})$,
where $\Omega$ is the cell volume.
RI have been computed~\cite{baroni86,umari01} through Eq.~(\ref{eq1}), obtaining
$\tensor{\bf A}^\nu$ by finite-differences derivation of
$\tensor{\bm{\epsilon}}^\infty$ with respect $u_{k\gamma}$.
This approach requires 36$N^{\rm at}$ linear response
calculations, where $N^{\rm at}$ is the number of atoms.
Thus, the scaling of the RI calculation is the same as that of the 
frequency calculation with a much larger prefactor.
This has limited the applications of this approach to small systems.
RI have also been computed from the 
the dynamical autocorrelation functions of
$\tensor{\bm{\epsilon}}^\infty$ in a molecular 
dynamics (MD) run~\cite{putrino02}. 
This approach also  copes with
liquids or anharmonic solids, but 
is very demanding, requiring the calculation of
$\tensor{\bm{\epsilon}}^\infty$ at each MD step. 

Alternatively, RI can be obtained knowing the second order derivative
of the DFT density matrix $\rho=\sum_v |\psi_v\rangle\langle\psi_v|$,
being $|\psi_v\rangle$ the normalized occupied Kohn-Sham (KS) eigenstates~\cite{DFT}.
In fact, according to the well known Hellmann-Feynman theorem
\[
\frac{\partial {\cal E}^{\rm el}}{\partial u_{k\gamma}}= 2\,
{\rm Tr}\left\{\rho\frac{\partial v^{\rm ext}}{\partial u_{k\gamma}}\right\},
\]
where ${\rm Tr}\{O\}$ is the trace of the operator $O$, and
$v^{\rm ext}$ is the external ionic potential (the KS self-consistent 
potential is
$V^{\rm KS}=V^{\rm Hxc}+v^{\rm ext}$, where $V^{\rm Hxc}$ is the sum of
the Hartree and the exchange-correlation potential).
Thus
\begin{equation}
\label{eq6}
\frac{\partial^3{\cal E}^{\rm el}}{\partial E_l \partial E_m\partial u_{k\gamma}}= 2\,
{\rm Tr}\left\{\left(\frac{\partial^2 \rho}{\partial E_l \partial E_m}
\right)\frac{\partial v^{\rm ext}}{\partial u_{k\gamma}}\right\}.
\end{equation}
The $\partial^2\rho/(\partial E_l \partial E_m)$ calculation
requires six second-order calculations, instead of the 
36$N^{\rm at}$ first-order calculations needed for the
finite-differentiation~\cite{umari01}. Because of this better size-scaling, the
$\tensor{\bf A}^\nu$ calculation
through Eq.~(\ref{eq6}) is much more efficient and
the time for RI calculation is negligible compared to that 
for $\omega_\nu$ in large systems.

The approach based on Eq.~(\ref{eq6}) has already been used in isolated 
molecules~\cite{frisch86} but never in extended systems.
Indeed, in solids the calculation of $\partial^2\!\rho/(\partial E_l \partial E_m)$ is not trivial because the position operator, required by  
the electric field perturbation,
is ill-defined in periodic boundary conditions.
Because of this, although a formalism to calculate 
derivatives of $\rho$ at any order was proposed by Gonze already in 1995
~\cite{gonze95}, only very recently Nunes and Gonze~\cite{nunes01} were able
to include perturbations due to macroscopic electric fields. To do that, they
use the polarization-Berry phase formalism~\cite{kingsmith93},
arguing that this concept remains valid in the presence of finite
electric fields. This approach has been applied so far to a
one dimensional non-self-consistent model~\cite{nunes01}.
In the following we give an expression for the second derivative of $\rho$,
that does not require the Berry phase formalism to cope with uniform electric 
fields, and we use it to compute $\tensor{\bf A}^\nu$
in real systems with a DFT self-consistent Hamiltonian.

The derivative of $\rho$ with respect to two
generic perturbation parameters $\lambda$ and $\mu$ is:
\begin{eqnarray}
\label{sec3}
\frac{\partial^2\rho}{\partial\lambda\partial\mu}&=&
\sum_v
\left(
|P\eta_v^{(\lambda,\mu)}\rangle
\langle\psi_v|+
|P\frac{\partial\psi_v}{\partial\lambda}\rangle
\langle \frac{\partial\psi_v}{\partial\mu}P|+
\right. \nonumber \\
& &
\left.
-\sum_{v'} |\psi_{v'}\rangle
\langle\frac{\partial\psi_{v'}}{\partial\lambda}P|P
\frac{\partial\psi_v}{\partial\mu}\rangle
\langle\psi_v|\right) + cc,
\end{eqnarray}
where $P=({\bf 1} -\rho)$ is the projector on the empty state subspace,
the sums over $v$ and $v'$ run over the occupied states,
and $|\eta_v^{(\lambda,\mu)}\rangle$ are the second derivatives of the
occupied KS-orbitals in the parallel-transport gauge~\cite{gonze95}.
According to our derivation:
\begin{eqnarray}
\label{eq30} 
|P\frac{\partial\psi_v}{\partial\lambda}\rangle&=&
\tilde G_v \left[ \frac{\partial V^{\rm KS}}{\partial \lambda},\rho
\right] |\psi_v\rangle, \\
\label{sec1}
|P\eta_v^{(\lambda,\mu)}\rangle &=&
\tilde G_v \left\{~
\frac{\partial^2V^{\rm KS}}{\partial\lambda\partial\mu}
~+~
\left[ \frac{\partial V^{\rm KS}}{\partial\lambda},
\frac{\partial\rho}{\partial\mu} \right]
\right.~+ \nonumber \\
&&
\left.
+~\left[ \frac{\partial V^{\rm KS}}{\partial\mu},
\frac{\partial\rho}{\partial\lambda} \right] 
~\right\}|\psi_v\rangle .
\end{eqnarray}
Here,
\[
\tilde G_v = \sum_c \frac{|\psi_c\rangle\langle\psi_c|}{\epsilon_v-\epsilon_c}
\]
is 
the  Green function operator projected on the empty states 
$|\psi_c\rangle$~\cite{note1},
$[A,B]=AB-BA$, and the first derivative of the density matrix is:
\begin{equation}
\label{eq31}
\frac{\partial \rho}{\partial \mu} = \sum_v 
|P\frac{\partial\psi_v}{\partial\mu}\rangle\langle\psi_v|+cc.
\end{equation}
Since $\partial V^{\rm KS}/\partial \lambda$ and 
$\partial^2 V^{\rm KS}/(\partial \lambda \partial \mu)$ depend on 
$\partial \rho/\partial \lambda$, $\partial \rho/\partial \mu$, and 
$\partial^2\! \rho/(\partial \lambda \partial \mu)$, Eqs.~(\ref{sec3}-\ref{eq31}), should be solved self-consistently.

The advantage of the present formulation, compared to that 
of Ref.~\cite{gonze95}, lies
in the introduction of the commutators of Eqs.~(\ref{eq30},\ref{sec1}). 
Thanks to the commutators, all the quantities needed with our formalism are well defined in an extended insulator,
even if the perturbation $\mu$ or $\lambda$ are the component $E_l$ of a
a uniform electric field, i.e. if $\partial V^{\rm KS}/\partial \lambda =
-e r_l + \partial V^{\rm Hxc}/\partial{E_l}$ ~\cite{nota2}, being $r_l$ the
$l^{th}$ Cartesian component of the position operator ${\bf r}$, and $e$ the
electron charge. In particular,  
in an insulator, the commutators $[{\bf r},\rho]$ and
$[{\bf r}, \partial\rho/\partial \mu]$ in Eqs.~(\ref{eq30},\ref{sec1}) 
are well
defined, bounded  operators, since the density matrix is
localized ($\langle {\bf r}'' |\rho| {\bf r}' \rangle$ goes
to zero exponentially for $|{\bf r}''-{\bf r}'|\rightarrow \infty$).

Finally, in a periodic system, the right-hand side of Eq.~(\ref{sec1}) can be easily
computed in terms of the $|u^{\bf k}_i\rangle$, that are the periodic parts 
of the Bloch-wavefunctions $|\psi^{\bf k}_i\rangle$ with reciprocal-lattice vector ${\bf k}$,
using the substitutions:
\begin{eqnarray}
\langle\psi^{\bf k}_c|\left[r_l,\rho\right]|
\psi^{\bf k}_v\rangle= i
\sum_{v'} \langle u^{\bf k}_c|
\frac{\partial |u^{\bf k}_{v'}\rangle
\langle u^{\bf k}_{v'}|}{\partial k_l} |u^{\bf k}_v\rangle \\
\langle\psi^{\bf k}_{c}|\left[r_l,\frac{\partial \rho}{\partial E_m}\right]|
\psi^{\bf k}_{v}\rangle= i\sum_{v'}
\langle u^{\bf k}_{c}|
\frac{\partial |P_{{\bf k}}\frac{\partial u^{\bf k}_{v'}}{\partial E_m}\rangle
\langle u^{\bf k}_{v'}|}{\partial k_l} |u^{\bf k}_{v}\rangle 
\label{trics}
\end{eqnarray}
where $l$ and $m$ are Cartesian indexes, $c$ is an empty band index,
$v$ and $v'$ are occupied band indexes,
and $P_{{\bf k}}$ is the projector on the empty subspace of the 
point ${\bf k}$.
In our implementation, the derivative with respect to $k_l$ in  the right-hand
side of Eq.~(\ref{trics}) is computed numerically by finite-differences,
using an expression independent from the arbitrary wavefunction-phase, as in Refs.~\cite{dalcorso94,nunes01}.

\begin{figure}
\centerline{\includegraphics[width=85mm]{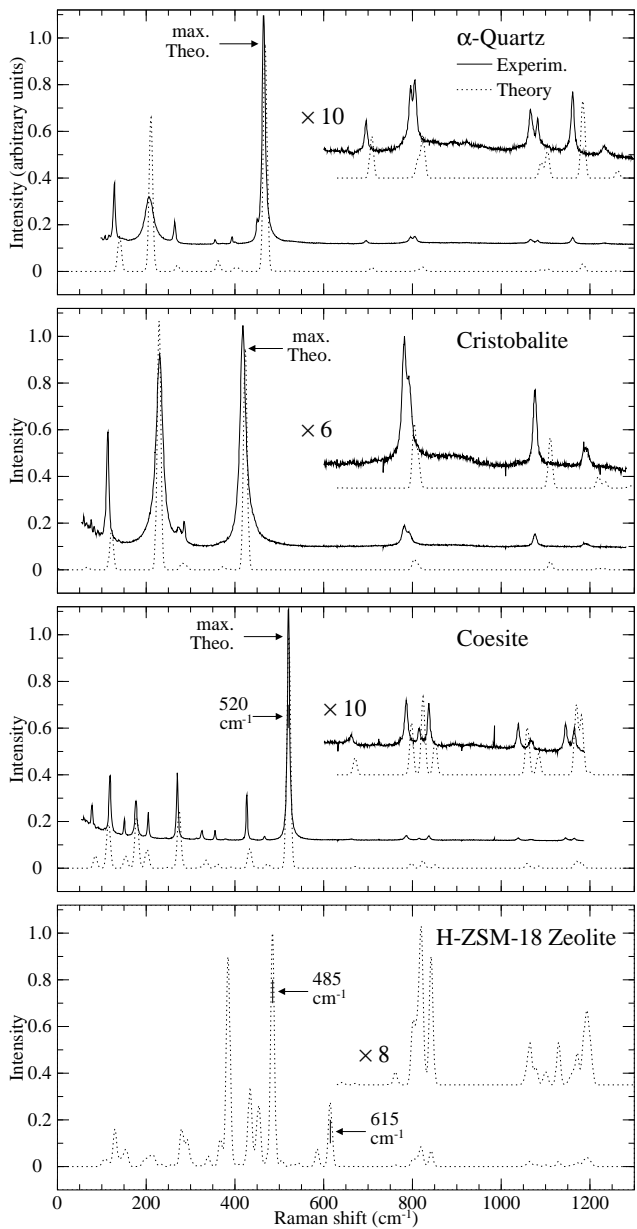}}
\caption{Vibrational Raman spectra of various SiO$_2$ polymorph powders.
Measurements are from Refs.~\protect\cite{coesite}.
Theoretical frequencies are rescaled by +5\%, and the spectra are convoluted
with a uniform Gaussian broadening having 4.0 cm$^{-1}$ width.}
\label{fig1}
\end{figure}

\begin{table}
\caption{Raman activity in Si computed with our approach 
($\gamma_{\rm SOR}$), and by finite differences ($\gamma_{\rm FD}$). $N$ is the number of inequivalent
k-points.  }
\label{tab1}
\begin{ruledtabular}
\begin{tabular}{c|rrrrrr}
           $N$&     2&    10&    28&    60&   110&   182\\
\hline
$\gamma_{\rm SOR}$&  8.54&  5.30&  5.32&  5.39&  5.40&  5.40\\
 $\gamma_{\rm FD}$& 18.99&  7.09&  5.69&  5.45&  5.41&  5.40\\
 $\gamma_{\rm FD}$  Ref.~\cite{baroni86}&      &7.10 & &      &      &      \\
\end{tabular}
\end{ruledtabular}
\end{table}

We test our approach on Si in the diamond phase, where 
the Raman activity is determined by 
$\gamma = a \partial\epsilon^{\infty}_{11}/\partial u$
~\cite{baroni86}, where $a= 10.20~a.u.$ is the lattice spacing
and $u$ the displacement of one atom
along the $(1,1,1)$ direction~\cite{technicalities}.
We compute $\gamma$ for various grids of k-points, using both 
our second order response method ($\gamma_{\rm SOR}$) and by finite
differentiation with respect to the atomic displacement ($\gamma_{\rm FD}$), Tab.~\ref{tab1}.
At convergence the two approaches are completely equivalent.

As a second application, we consider 
tetrahedral SiO$_2$. In this class of materials, that includes
the all-silica zeolites, the quartz, cristobalite,
tridymite and coesite polymorphs of SiO$_2$, and vitreous silica (v-SiO$_2$),
each Si atom is tetrahedrally
coordinated to four O atoms and each O atom is bonded to two Si atoms.
The properties of these systems can be 
effectively described in terms of the $n$-membered rings ($n$-MRs) of 
tetrahedra contained in their structure~\cite{sharma81,hosono01,pasquarello98}.
E.g., a clear correlation between
the presence of 3- and 4-MRs and the degradation of optical v-SiO$_2$ fibers
under UV radiation has been observed~\cite{hosono01}.
In the v-SiO$_2$ Raman spectra
the two sharp peaks at 490 cm$^{-1}$ ($D_1$ line)  and 606 cm$^{-1}$
($D_2$ line), have been attributed to the
breathing mode (BM) of the O atoms towards the ring center 
of 4-MRs and 3-MRs, respectively~\cite{sharma81}.
This attribution has been confirmed by DFT vibrational frequency 
calculations~\cite{pasquarello98}.
The attribution
would be further supported by experimental measurements on well characterized
crystalline polymorphs containing 3- and 4-MRs.
However,
the strong Raman peak at 520 cm$^{-1}$ in coesite, a phase that contains  4-MRs,  is 
shifted by 30
cm$^{-1}$ with respect to the $D_1$ line in v-SiO$_2$, and no Raman 
measurements has been published
on the  H-ZSM-18 zeolite, that is the only known SiO$_2$ crystalline polymorph with
3-MRs~\cite{zeolite}. Interestingly this zeolite contains 4-MRs as well.

\begin{figure}[t]
\centerline{\includegraphics[width=85mm]{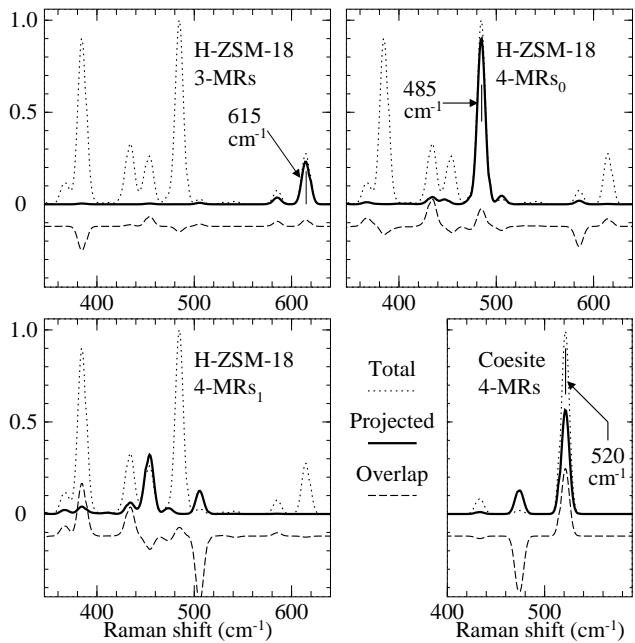}}
\caption{Raman intensities projected on the breathing
modes of various rings labeled 3-MRs, and 4-MRs$_x$ (see the text).
For clarity, the overlap intensity ($I^\nu_{\rm overlap}$ in the text)
is shifted vertically.}
\label{fig2}
\end{figure}

To clarify this topic,
we compute the Raman spectra of  $\alpha$-quartz, coesite, 
$\alpha$-cristobalite, and H-ZSM-18~\cite{technicalities,technicalities2}.
In Fig.~\ref{fig1}, we compare our results with the 
available experimental spectra~\cite{coesite}. The vibrational
frequencies are systematically underestimated by 5\% by our calculation.
To simplify the comparison with the experiments, in  Fig.~\ref{fig1} and \ref{fig2},
the theoretical frequencies are
multiplied by a scaling factor of 1.05.
The ability of the method in reproducing quantitatively
all the measured features is evident.

In order to associate Raman peaks of Fig.~\ref{fig1} to the small-ring BMs,
we project the vibrational eigenmode                   
${\bf w}^\nu$
on the subspace generated by the BMs of a given kind of rings, $\cal R$, and on
the corresponding complementary subspace, $\bf  \bar{\cal R}$.
We use the two resulting projected vectors to decompose
$\tensor {\bf A}^\nu$ so that
$\tensor {\bf A}^\nu=\tensor {\bf A}^\nu_{\cal R}+\tensor {\bf A}^\nu_{\bf \bar{\cal R}}$.  
Since $I^\nu$
is quadratic in $\tensor{\bf A}^\nu$, see Eq.~(\ref{eq1}), $I^\nu=
I^\nu_{\cal R} + I^\nu_{\bf \bar{\cal R}}+ I^\nu_{\rm overlap}$, where 
$I^\nu_{\rm overlap}$ is the term bilinear in 
$\tensor {\bf A}^\nu_{\cal R}$ and  $\tensor {\bf A}^\nu_{\bf \bar{\cal R}}$.
A Raman peak can be associated to a ring BM (i.e. the Raman activity is solely
due to the BM ) if, and only if,
$I^\nu_{\bf {\cal R}}\gg |I^\nu_{\rm overlap}|$.

The structure of H-ZSM-18~\cite{zeolite}
contains two equivalent 3-MRs and two kinds of 4-MRs which we will
call 4-MRs$_0$, and 4-MRs$_1$~\cite{rings}.
In Fig. ~\ref{fig2}, we show the projected Raman spectra of the zeolite and the
coesite.
In the H-ZSM-18 spectrum, the peaks at 485, and 615
cm$^{-1}$ are very well described by the BM of 4-MRs$_0$ and
3-MRs, respectively.
A direct analysis of the vibrational eigenmodes shows that
both BMs are decoupled from other modes.
The frequencies of the two peaks are very close to those of the measured
$D_1$ and $D_2$ lines in v-SiO$_2$ (490, and 606 cm$^{-1}$), thus confirming
that these lines are due to rings BMs~\cite{sharma81,pasquarello98}.
However, the presence of small-MRs in a structure does not guarantee,
in general, the occurrence of completely decoupled BMs.
This is the case of the 4-MRs in coesite and
the 4-MRs$_1$ in the zeolite, whose BMs exhibit a large 
$|I^\nu_{\rm overlap}|$, see Fig. ~\ref{fig2}.
These overlaps imply the existence of a coupling with other modes,
that, in turn, explains the 30 cm$^{-1}$ difference
between the 4-MRs frequency of coesite and that of the $D_1$ line of v-SiO$_2$.
A comparable frequency shift from the  $D_1$ line is observed, 
with opposite sign, for the 4-MRs$_1$ BMs in the zeolite.

In conclusion, with the aim of building an instrument for the routine
interpretation of Raman spectra, we developed a method for
the efficient calculation of Raman intensity.
We computed the Raman spectra of
SiO$_2$ polymorphs containing up to 102 atoms.
We found that: i) not all the small-membered-rings have decoupled 
breathing modes,
ii) the H-ZSM-18 zeolite provides 
decoupled breathing mode of 4- and 3-membered rings, whose frequencies nicely coincide with 
the $D_1$ and $D_2$ lines of vitreous silica. An experimental 
determination of the Raman spectra of this zeolite can thus provide an
experimental calibration for the determination of the density of 
decoupled small membered rings in vitreous silica. 

Calculations were performed at IDRIS supercomputing center.
Our approach was implemented in the PWSCF code~\cite{pwscf}.



\begin{references}
\bibitem{gonze95}
X.\ Gonze, Phys.\ Rev.\ A {\bf 52}, 1096 (1995).
Note that
the term depending on the Lagrange multipliers should not be omitted
in the derivation of Eq.~(55) from Eq.~(54).
\bibitem{note1}
The operator $\tilde G_v$ can be efficiently applied to any vector
by solving a linear system, following Refs.~\cite{DFPT,gonze95}.
\bibitem{technicalities2}
We use the experimental lattice constants~\cite{str,zeolite},
1 special k-point for H-ZSM-18, and  3 k-points 
in the other systems.
Internal coordinates are relaxed by energy minimization.
Raman spectra are computed in backscattering geometry taking into account the TO-LO
splitting of the frequencies~\cite{DFPT}.
\bibitem{rings}
In a unit cell, there are twelve equivalent 4-MRs$_0$, which
are jointed in four groups of three. In each group the three rings
are sharing one Si atom.
There are six equivalent 4-MRs$_1$ which are disjointed between them.
\end{references}
\end{document}